\documentstyle[aps,twocolumn,epsf]{revtex}

\begin{document}
\title{
{\em Invited Presentation at the NATO Advanced Research Workshop on
Symmetry and Pairing in Superconductors, Yalta (Crimea),
April 28--May 2, 1998 (to be published in NATO ASI Series)}\\
\vskip 2cm
Probing mixed $s\pm id$ pairing state via thermoelectric response
of $SND$ junction}
\author{Sergei A. Sergeenkov\cite{byline}}
\address{Bogoliubov Laboratory of Theoretical Physics,
Joint Institute for Nuclear Research, 141980 Dubna, Moscow Region, Russia}
\address{(\today)}
\address{~}
\address{
\centering{
\medskip\em
\begin{minipage}{14cm}
{}~~~
The thermoelectric response of $SND$ configuration is
considered within the generalized Ginzburg-Landau theory for a
homogeneous admixture of $s$-wave and $d$-wave superconductors. The resulting
thermopower $\Delta S(T,\theta ,t_c)=S_p(\theta ,t_c)-
B(\theta ,t_c)(T_{cs}-T)$ is found
to strongly depend on the relative phase $\theta =\phi _s-\phi _d$
between the two superconductors ($t_c\equiv T_{cd}/T_{cs}$ where $T_{cs}$
and $T_{cd}$ are the corresponding critical temperatures).
Two independent mechanisms are shown to
contribute to the peak value $S_p(\theta ,t_c)$. One, based on
the charge imbalance between the quasiparticles and Cooper
pairs (described by the corresponding chemical potentials,
$\mu _q$ and $\mu _p$) due to the normal metal insert, results in a
pronounced maximum of the peak near $\theta =\pm \pi /2$
(the so-called $s\pm id$ mixed pairing state) for two identical
superconductors with $T_{cd}=T_{cs}$. This mechanism can be realized
in a $d$-wave orthorhombic sample
(like $YBCO$) with twin boundaries which are represented by tetragonal
regions of variable width, with a reduced chemical potential.
Another mechanism (not related to the charge imbalance effects) occurs when
two different superconductors with $T_{cd}\neq T_{cs}$ are used
for $SND$ junction. It gives rise to $S_p(\theta ,t_c)\propto
1-t_c$ and can be realized via the junction comprising an $s$-wave low-$T_c$
superconductor (like $Pb$) and a $d$-wave high-$T_c$ superconductor
(like orthorhombic $YBCO$).
The experimental conditions (based on the previous experience with $SNS$
junctions) under which the predicted
behavior of the induced differential thermopower can be measured are
discussed.
{}~\\
\end{minipage}
}}

\maketitle

\narrowtext

\section{Introduction}
During the last few years the order parameter symmetry has been
one of the intensively debated issues in the field of high-$T_c$
superconductivity (HTS). A number of experiments points to its
$d_{x^2-y^2}$-wave character~\cite{1}. Such an unconventional
symmetry of the order parameter has also important implications for the
Josephson physics
because for a $d$-wave superconductor the Josephson coupling is subject
to an additional phase dependence caused by the internal phase
structure of the wave function.  The phase properties of
the Josephson effect have been discussed within the framework of the
generalized Ginzburg-Landau (GL)~\cite{2} as well as the tunneling
Hamiltonian approach~\cite{3}. It was found~\cite{4} that the
current-phase
relationship depends on the mutual orientation of the two coupled
superconductors and their interface.  This property is the basis of
all the phase sensitive experiments probing the order parameter symmetry.
In particular, it is possible to create multiply connected $d$-wave
superconductors which generate half-integer flux quanta as observed in
experiments~\cite{5}.
Various interesting phenomena occur in interfaces of $d$-wave
superconductors. For example, for an interface to a normal metal a bound
state appears at zero energy giving rise to a zero-bias anomaly in the
$I$-$V$-characteristics of quasiparticle tunneling~\cite{6,7}
while in such an interface to an $s$-wave superconductor the energy minimum
corresponds to a Josephson phase different from $0$ or $\pi$.
By symmetry, a small $s$-wave component always coexists with a predominantly
$d$-wave order parameter in an orthorhombic superconductor such as $YBCO$,
and changes its sign across a twin boundary~\cite{8}.
Besides, the $s$-wave and $d$-wave order parameters can form a complex
combination, the so-called $s\pm id$-state which is
characterized by a local breakdown of time reversal symmetry
${\cal T}$ either near surfaces~\cite{9,10,11,12} or near the twin
boundaries represented by tetragonal regions with a reduced
chemical potential~\cite{13}. Both scenarios lead to
a phase difference of $\pm \pi/2$, which corresponds to two
degenerate states~\cite{14,15}. Moreover,
the relative phase oscillations between two condensates with
different order parameter symmetries could manifest themselves through
the specific collective excitations ("phasons")~\cite{16}.

At the same time, a rather sensitive differential technique to probe
sample inhomogeneity for temperatures just below $T_c$, where phase
slippage events play an important role in transport characteristics
has been proposed~\cite{17} and successfully applied~\cite{18} for
detecting small changes in thermopower of a specimen due to the
deliberate insertion of a macroscopic $SNS$ junction made of a normal-metal
layer $N$, used to force pair breaking of the superconducting
component when it flows down the temperature gradient.
In particular, a carrier-type-dependent thermoelectric response of such
a $SNS$ configuration in a $C$-shaped $Bi_xPb_{1-x}Sr_2CaCu_2O_y$ sample
has been registered and its $\Lambda$-shaped temperature behavior
around $T_c$ has been explained within the framework of GL
theory~\cite{18,19}.

In the present paper, we consider theoretically the case of $SND$ junction
and discuss its possible implications for the above-mentioned type of
experiments. The paper is organized as follows. In Section II
we briefly review the experimental results for $SNS$ configuration (with
both holelike and electronlike carriers of the normal-metal $N$ insert)
and present a theoretical interpretation of these results,
based on GL free energy functional. The crucial role of the
difference between the quasiparticle $\mu _q$ and pair $\mu _p$ chemical
potentials in understanding the observed phenomena is emphasized.
In Section III, extending the early suggested~\cite{10,13} GL theory of
an admixture of $s$-wave and $d$-wave superconductors by
taking into account pair-breaking effects with $\mu _q\neq \mu _p$, we
calculate the differential thermopower
$\Delta S$ of $SND$ configuration near $T_c$. The main
theoretical result of this Section is prediction of a rather specific
dependence of $\Delta S$ on relative phase shift $\theta  =\phi _s-\phi _d$
between the two superconductors.
Two independent mechanisms contributing to the peak value of the thermopower
are discussed. One, based on
the charge imbalance between the quasiparticles and Cooper
pairs due to the normal metal insert, is discussed in Section IIIA.
It results in a
pronounced maximum of the peak near $\theta =\pm \pi /2$
(the so-called $s\pm id$ mixed pairing state) for two identical
superconductors with $T_{cd}=T_{cs}\equiv T_c$. This mechanism can be
realized, e.g., in a $d$-wave orthorhombic sample
(like $YBCO$) with twin boundaries which are represented by tetragonal
regions of variable width, with a reduced chemical potential.
Another mechanism (not related to the charge imbalance effects), discussed
in Section IIIB, occurs when
two different superconductors with $T_{cd}\neq T_{cs}$ are used
for $SND$ junction. This situation can be realized for an $s$-wave low-$T_c$
superconductor (like $Pb$) and a $d$-wave high-$T_c$ superconductor
(like orthorhombic $YBCO$).

\section{$SNS$ configuration: a review}

\subsection{Experimental setup and main results}

Before turning to the main subject of the present paper, let us
briefly review the previous results concerning a carrier-type-dependent
thermoelectric response of $SNS$ configuration in a $C$-shaped
$Bi_xPb_{1-x}Sr_2CaCu_2O_y$ sample (see Ref.~\cite{18} for details).
The sample geometry used is sketched in Fig.1, where the contact
arrangement and
the position of the sample with respect to the temperature gradient
$\nabla _xT$ is shown as well.
Two cuts are inserted at $90^o$ to each other
into a ring-shaped superconducting sample. The first cut lies parallel
to the applied temperature gradient serving to define a vertical symmetry
axis. The second cut lies in the middle of the right wing, normal to the
symmetry axis, separating an $s$-wave superconductor ($S'=S$) from another
$s$-wave superconductor ($S''=S$) and completely interrupting the passage
of supercurrents in
this wing. The passage of any normal component of current density is
made possible by filling up the cut with a normal metal $N$.
The carrier type of the normal-metal insert
$N$ was chosen to be either an electronlike $N_e$ (silver) or holelike $N_h$
(indium).
Thermal voltages resulting from the same temperature gradient acting on
both continuous and normal-metal-filled halves of the sample were detected
as a function of temperature around $T_c$. The measured difference between
the thermopowers of the two halves $\Delta S=S_R-S_L$ was found to
approximately follow the linear dependence
\begin{equation}
\Delta S(T)\simeq S_p-B(T_c-T),
\end{equation}
where $S_p=\Delta S(T_c)$ is the peak value of $\Delta S(T)$ at $T=T_c$,
and $B$ is a constant. The best fit of the experimental data with the
above equation yields the following values for silver (Ag) and indium (In)
inserts, respectively: (i) $S_p(Ag)=-0.26\pm 0.01 \mu V/K$, $B(Ag)=
-0.16\pm 0.1 \mu V/K^2$; (ii) $S_p(In)=0.83\pm 0.01 \mu V/K$, $B(In)=
0.17\pm 0.1 \mu V/K^2$.

\begin{figure}[htb]
\epsfxsize=7.5cm
\centerline{\epsffile{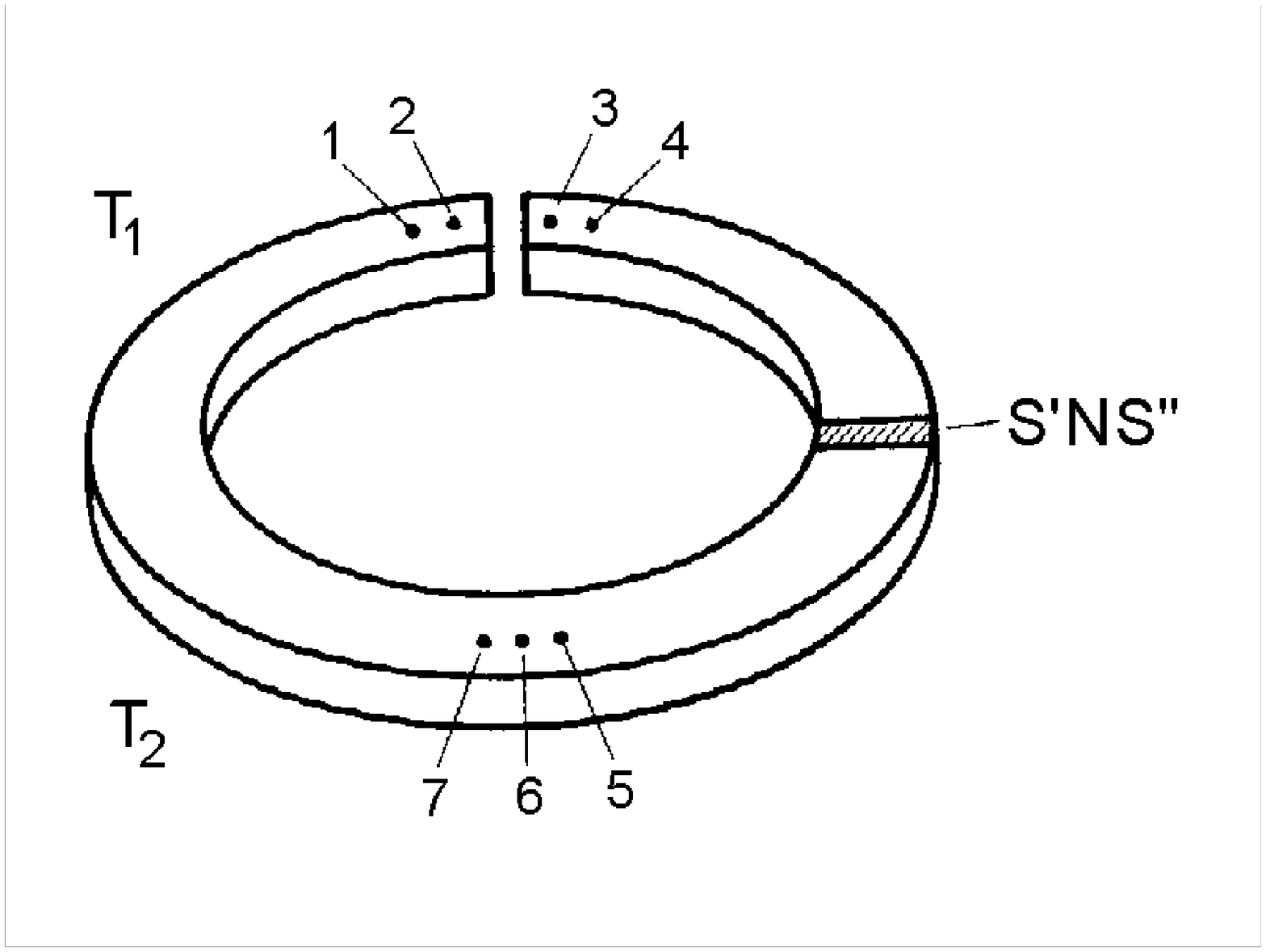} }
\caption{Schematic view of the sample geometry with $S'NS''$-junction and
contacts configuration. The thermopowers $S_R$ and $S_L$ result from the
thermal voltages detected by the contact pairs $4-5$ and $1-7$, respectively.}
\end{figure}

\subsection{Interpretation}

Assuming that the net result of the normal-metal insert is to break up
Cooper pairs that flow toward the hotter end of the sample and to produce
holelike (In) or electronlike (Ag) quasiparticles, we can write the
difference in the generalized GL free energy functional $\Delta {\cal G}$ of
the right and left halves of the $C$-shaped sample as
\begin{equation}
\Delta {\cal G} [\psi ]=\Delta {\cal F}[\psi ]-\Delta \mu |\psi |^2,
\end{equation}
where
\begin{equation}
\Delta {\cal F}[\psi ]\equiv {\cal F}_R-{\cal F}_L=a(T)|\psi |^2+
\frac{\beta}{2}|\psi |^4
\end{equation}
and
\begin{equation}
\Delta \mu \equiv \mu _R-\mu _L.
\end{equation}
Here $\psi =|\psi |e^{i\phi}$ is the superconducting order parameter,
$\mu _p$ and $\mu _q$ are the chemical potentials of quasiparticles and
Cooper pairs, respectively; $a(T)=\alpha (T-T_c)$ and the GL parameters
$\alpha$ and $\beta$ are related to the critical temperature $T_c$,
zero-temperature gap $\Delta _0=1.76k_BT_c$, the Fermi energy $E_F$, and
the total particle number density $n$ as $\alpha =2\Delta _0k_B/E_F$ and
$\beta =\alpha T_c/n$.

As usual, the equilibrium state of such a system is determined from the
minimum energy condition $\partial {\cal G}/\partial |\psi |=0$ which
yields for $T<T_c$
\begin{equation}
|\psi _0|^2=\frac{\alpha (T_c-T)+\Delta \mu}{\beta}
\end{equation}
Substituting $|\psi _0|^2$ into Eq.(2) we obtain for the generalized free
energy density
\begin{equation}
\Delta \Omega (T)\equiv \Delta {\cal G} [\psi _0]=-
\frac{[\alpha (T_c-T)+\Delta \mu ]^2}{2\beta}
\end{equation}
In turn, the observed difference of thermopowers $\Delta S(T)$ can be
related to the
corresponding difference of transport entropies $\Delta \sigma \equiv
-\partial \Delta \Omega /\partial T$ as $\Delta S(T)=\Delta \sigma (T)/nq$,
where $q$ is the charge of the quasiparticle.
Thus finally the thermopower associated with a pair-breaking
event reads
\begin{equation}
\Delta S(T)=\frac{\Delta \mu}{qT_c}-\frac{2\Delta _0k_B}
{q\tilde E_F T_c}(T_c-T),
\end{equation}
where $\tilde E_F=E_F-\mu _q$ accounts for the shift of the Fermi energy
$E_F$ due to the quasiparticle chemical potential $\mu _q$.
Let us discuss now separately the case of $In$ and $Ag$ normal-metal inserts.

\subsubsection{$N=In$ (holelike metal insert)}

In this case, the principal carriers are holes, therefore $q=+e$ in Eq.(7).
Let the holelike quasiparticle chemical potential (measured relative to
the Fermi level of the free-hole gas) be positive, then $\mu _q=+\mu$
and $\Delta \mu \equiv \mu _q-\mu _p=\mu -(-2\mu)=3\mu$. Here $\mu _p=-2\mu$
comes from the change of the pair chemical potential of the holelike
condensate with respect to the holelike quasiparticle branch. Therefore,
for this case Eq.(7) takes the form
\begin{equation}
\Delta S^h(T)=
3\left (\frac{k_B}{e}\right )\left (\frac{\mu}{k_BT_c}\right )-
\frac{2\Delta _0k_B}{e\tilde E_F^h T_c}(T_c-T),
\end{equation}
where $\tilde E_F^h=E_F-\mu$.

\subsubsection{$N=Ag$ (electronlike metal insert)}

The principal carriers in this case are electrons, therefore $q=-e$. The
electronlike quasiparticle chemical potential (measured relative to the
Fermi level of the free-hole gas) is $-\mu$. Then $\mu _q=-\mu$ and
$\Delta \mu =-\mu -(-2\mu )=\mu$. For this case Eq.(7) takes the form
\begin{equation}
\Delta S^e(T)=
-\left (\frac{k_B}{e}\right )\left (\frac{\mu}{k_BT_c}\right )+
\frac{2\Delta _0k_B}{e\tilde E_F^e T_c}(T_c-T),
\end{equation}
where $\tilde E_F^e=E_F+\mu$.

Using the above-mentioned experimental findings for the slope $B$ and the
peak $S_p$ values for the two normal-metal inserts (see Eq.(1)), we can
estimate the order of magnitude of the Fermi energy $E_F$ and quasiparticle
potential $\mu$. The result is: $E_F=0.16eV$ and $\mu =5\times 10^{-3}eV$,
in reasonable agreement with the other known estimates of these parameters.
Besides, as it follows from Eqs.(8) and (9), the calculated ratio for peaks
$|S_p(In)/S_p(Ag)|=3$ is very close to the corresponding experimental value
$|S_p^{exp}(In)/S_p^{exp}(Ag)|=3.2\pm 0.2$ observed by Gridin et
al~\cite{18}.

\section{$SND$ configuration: prediction}

Since Eqs.(2)-(4) do not depend on the phase of the order parameter, they
will preserve their form for a $DND$ junction (created by two $d$-wave
superconductors, $S'=S''=D$, see Fig.1) bringing about
the result similar to that given by Eqs.(7)-(9). It means that the
experimental method
under discussion (and its interpretation) can not be used to tell
the difference between $SNS$ and $DND$ configurations, at least for
temperatures close to $T_c$. As for low enough temperatures, the situation
may change drastically due to a markedly different behavior of $s$-wave and
$d$-wave order parameters at $T\ll T_c$. As we will show, this method,
however,
is quite sensitive to the mixed $SND$ configuration (when $S'=S$
has an $s$-wave symmetry while $S''=D$ is of a $d$-wave symmetry type, see
Fig.1) and predicts a
rather specific relative phase ($\theta =\phi _s-\phi _d$) dependences of
both the slope $B(\theta )$ and peak $S_p(\theta )$ of the observable
thermopower difference $\Delta S(T,\theta )$.

Following Feder et al~\cite{13}, who incorporated chemical potential effects
near twin boundaries into the approach suggested by Sigrist et al~\cite{10},
we can represent
the generalized GL free energy functional $\Delta {\cal G}$ for $SND$
configuration of the $C$-shaped sample in the following form
\begin{equation}
\Delta {\cal G} [\psi _s,\psi _d]=\Delta {\cal G} [\psi _s]+
\Delta {\cal G} [\psi _d]+\Delta {\cal G}_{int},
\end{equation}
where
\begin{equation}
\Delta {\cal G} [\psi _{s}]=\Delta {\cal F}[\psi _{s}]-
\Delta \mu |\psi _{s}|^2,
\end{equation}
\begin{equation}
\Delta {\cal G} [\psi _{d}]=\Delta {\cal F}[\psi _{d}]-
\Delta \mu |\psi _{d}|^2,
\end{equation}
and
\begin{eqnarray}
\Delta {\cal G}_{int}=\gamma _1|\psi _s|^2|\psi _d|^2+
\frac{\gamma _2}{2}(\psi _s^{*2}\psi _d^2+\psi _s^2\psi _d^{*2})&&\\ \nonumber
-2\delta _1|\psi _s||\psi _d|-
\delta _2(\psi _s^{*}\psi _d+\psi _s\psi _d^{*}).&&
\end{eqnarray}
Here $\psi _n=|\psi _n|e^{i\phi _n}$ is the $n$-wave order parameter
($n=\{s,d\}$); $\Delta {\cal F}[\psi _{s,d}]$ is given by Eq.(3) with
the corresponding parameters $a_s(T)=\alpha _s(T-T_{cs}), \beta _s$,
$a_d(T)=\alpha _d(T-T_{cd})$, and $\beta _d$ for
$s$-wave and $d$-wave symmetry, respectively.

An equilibrium state of such a mixed system is determined from the
minimum energy conditions $\partial {\cal G}/\partial |\psi _s|=0$
and $\partial {\cal G}/\partial |\psi _d|=0$ which result in the following
system of equations for the two equilibrium order parameters
$\psi _{s0}$ and $\psi _{d0}$
\begin{eqnarray}
&&A_s(T)|\psi _{s0}|+\beta _s|\psi _{s0}|^3+
\Gamma (\theta )|\psi _{s0}||\psi _{d0}|^2=\Delta (\theta )|\psi _{d0}|\\
&&A_d(T)|\psi _{d0}|+\beta _d|\psi _{d0}|^3+
\Gamma (\theta )|\psi _{d0}||\psi _{s0}|^2=\Delta (\theta )|\psi _{s0}|
\end{eqnarray}
where $A_s(T)=a_s(T)-\Delta \mu$, $A_d(T)=a_d(T)-\Delta \mu$, and
we introduced relative phase $\theta =\phi _s-\phi _d$ dependent
parameters
\begin{eqnarray}
&&\Gamma (\theta )=\gamma _1+\gamma _2\cos 2\theta \\ \nonumber
&&\Delta (\theta )=\delta _1+\delta _2\cos \theta
\end{eqnarray}
Notice that the $\Delta (\theta )$ term favors $\theta =l\pi $ ($l$ integer),
while the $\Gamma (\theta )$ term favors $\theta =n\pi /2$ ($n=1,3,5\ldots$)
which corresponds to a ${\cal T}$-violating phase~\cite{13}.
In principle, we can resolve the above system (given by Eqs.(14)-(16)) and
find $\psi _{n0}$ for arbitrary set of parameters $\alpha _n$,
$\beta _n$, and $T_{cn}$. For simplicity, in what follows
we restrict our consideration to the two limiting cases which are of
the most importance for potential applications.

\subsection{Twin boundaries in orthorhombic $d$-wave superconductors}

First, let us consider the case of similar superconductors comprising the
$SND$ junction with $|\psi _{s0}|=
|\psi _{d0}|\equiv |\psi _0|$, $\alpha _s=\alpha _d\equiv \alpha$,
$\beta _s=\beta _d\equiv \beta$, and $T_{cs}=T_{cd}\equiv T_c$.
This situation is realized, for example, in a $d$-wave orthorhombic sample
(like $YBCO$) with twin boundaries which are represented by tetragonal
regions of variable width, with a reduced chemical potential~\cite{13}.
In this particular case, Eqs.(14) and (15) yield for $T<T_c$
\begin{equation}
|\psi _0|^2=\frac{\alpha (T_c-T)+\Delta \mu +\Delta (\theta )}
{\beta +\Gamma (\theta )}
\end{equation}
After substituting the thus found $|\psi _0|$ into Eq.(10) we obtain for the
generalized equilibrium free energy density
\begin{equation}
\Delta \Omega (T,\theta )\equiv \Delta {\cal G} [\psi _0]=-
\frac{[\alpha (T_c-T)+\Delta \mu +\Delta (\theta) ]^2}{\beta +
\Gamma (\theta )}
\end{equation}
which in turn results in the following expression for the thermopower
difference in a $C$-shaped sample with $SND$ junction (see Fig.1)
\begin{equation}
\Delta S(T,\theta )=S_p(\theta )-B(\theta )(T_c-T),
\end{equation}
where
\begin{equation}
S_p(\theta )=-\frac{2\Delta \mu}{qT_c}\left (\frac{1+
\tilde {\delta} \cos \theta}{1+\tilde {\gamma} \cos 2\theta}\right)
\end{equation}
and
\begin{equation}
B(\theta )=\frac{4\Delta _0k_B}{q\tilde {E_F}T_c}\left (\frac{1}{1+
\tilde {\gamma} \cos 2\theta}\right )
\end{equation}
Here, $\tilde \gamma =\gamma _2/(\gamma _1+\beta )$ and $\tilde \delta =
\delta _2/(\delta _1+\Delta \mu )$ with $\Delta \mu$ and $\beta$ defined
earlier.
\begin{figure}[htb]
\epsfxsize=8.5cm
\centerline{\epsffile{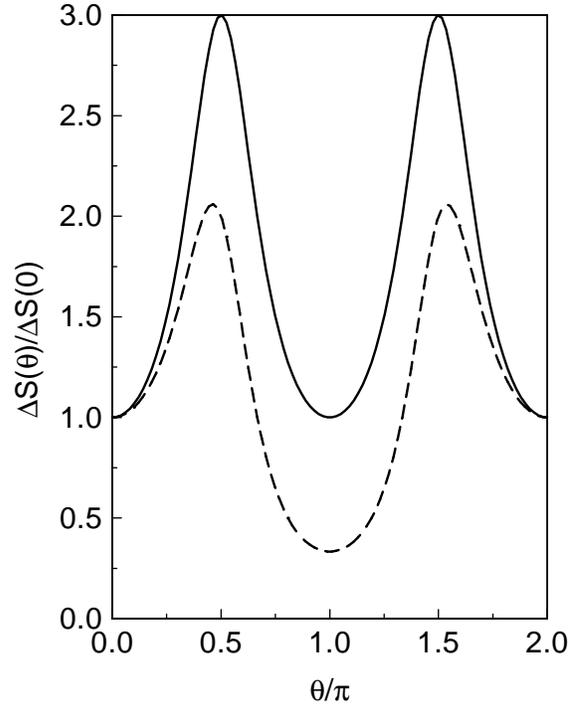} }
\caption{Predicted phase-dependent thermopower response of $SND$
configuration in a $C$-shaped sample (see Fig.1). Solid and dashed lines
depict, respectively, the relative phase $\theta$ dependence of the
normalized slope
$B(\theta )/B(0)$ and peak value $S_p(\theta )/S_p(0)$ of the induced
thermopower difference, according to Eqs.(20) and (21) with
$\tilde \gamma =\tilde \delta =1/2$.}
\end{figure}

Fig.2 shows the predicted $\theta$-dependent behavior of the normalized
slope $B(\theta )/B(0)$ (solid line) and the peak $S_p(\theta )/S_p(0)$
(dashed line) of the $SND$-induced thermopower
difference $\Delta S(T,\theta )$ near $T_c$, for $\tilde \gamma =
\tilde \delta =1/2$. As is seen, both the slope
and the peak exhibit a maximum for the $s\pm id$ state (at $\theta =\pi /2$)
and a minimum for the $s-d$ state (at $\theta =\pi$). Such sharp
dependencies suggest quite
an optimistic possibility to observe the above-predicted behavior of the
induced thermopower, using the described in Section II sample geometry
and experimental technique. Besides, by a controllable
variation of the carrier type of the normal-metal insert, we can get a more
detailed information about the mixed states and use it to estimate the
phenomenological parameters $\gamma _{1,2}$ and $\delta _{1,2}$.

\subsection{Low-$T_c$ $s$-wave superconductor and high-$T_c$ $d$-wave
superconductor}

Let us turn now to another limiting case when the two superconductors of
the $SND$ junction are different, so that
$|\psi _{s0}|\neq |\psi _{d0}|$, $\alpha _s\neq \alpha _d$,
$\beta _s\neq \beta _d$, and $T_{cs}\neq T_{cd}$ but the charge imbalance
effects are rather small and can be safely neglected, that is we assume
$\Delta \mu =0$, and $\Delta(\theta )=0$. Such a situation can be realized
for an $s$-wave low-$T_c$ superconductor (like $Pb$) and a $d$-wave
high-$T_c$ superconductor (like orthorhombic $YBCO$)~\cite{1}.
In fact, the solution for this particular case is well-known. It has been
discussed by Sigrist et al~\cite{10} in a somewhat different context.
The corresponding expressions for the equilibrium order parameters read
\begin{equation}
|\psi _{s0}|^2=\frac{\beta _da_s(T)-\Gamma (\theta )a_d(T)}
{\Gamma ^2(\theta )-\beta _s\beta _d},
\end{equation}
\vskip 0.5cm
\begin{equation}
|\psi _{d0}|^2=\frac{\beta _sa_d(T)-\Gamma (\theta )a_s(T)}
{\Gamma ^2(\theta )-\beta _s\beta _d},
\end{equation}
where $a_s(T)=\alpha _s(T-T_{cs})$ and $a_d(T)=\alpha _d(T-T_{cd})$.

After substituting this solution into Eq.(10) we obtain for the
the thermopower difference
\begin{equation}
\Delta S(T,\theta ,t_c)=S_p(\theta ,t_c)-B(\theta ,t_c)(T_{cs}-T),
\end{equation}
where $t_c=T_{cd}/T_{cs}$ and both the peak
\begin{equation}
S_p(\theta ,t_c)=(1-t_c)f(\theta ,t_c)
\end{equation}
and the slope
\begin{equation}
B(\theta ,t_c)=2\frac{f(\theta ,t_c)}{T_{cs}}
\end{equation}
are governed by a universal function
\begin{equation}
f(\theta ,t_c)=\frac{2\Delta _{s0}k_B\beta _st_c}{qE_F[\Gamma (\theta )
+\beta _st_c]}.
\end{equation}
Notice that in this case (when changes in chemical potentials can be
neglected) the peak's amplitude $S_p(\theta ,t_c)$ will be entirely dominated
by the critical temperatures difference $T_{cd}-T_{cs}$ of the two
superconductors because $f(\theta ,t_c)$ is a smooth function of $t_c$.
It would be interesting to test the predicted behavior of the induced
thermopower in a $C$-shaped sample with an $SND$ junction (see Fig.1) using
a low-$T_c$ $s$-wave and a high-$T_c$ $d$-wave superconductors.

In summary, to probe into the mixed $s\pm id$ pairing state of high-$T_c$
superconductors, we calculated the differential thermopower $\Delta S$ of
$SND$ junction in the
presence of the strong charge imbalance effects (due to a nonzero difference
between the quasiparticle $\mu _q$ and Cooper pair $\mu _p$ chemical
potentials) using the generalized Ginzburg-Landau theory for a
homogeneous admixture of $s$-wave and $d$-wave superconductors near $T_c$.
The calculated thermopower was found to strongly depend on the relative
phase $\theta =\phi _s-\phi _d$ between the two superconductors
exhibiting a pronounced maximum near the mixed $s\pm id$ state with
$\theta =\pm \pi /2$. The experimental conditions under which the predicted
behavior of the induced thermopower could be observed were discussed.

\end{document}